\documentclass[5p]{elsarticle}

\usepackage{hyperref}
\usepackage{epsfig, bm, amsfonts, amssymb}
\usepackage[normalem]{ulem}
\RequirePackage{color}

 \definecolor{MyDarkGreen}{rgb}{0.02,0.60,0.06}

\usepackage{epstopdf}

\journal{Physics Letters A}

\begin{document}

\begin{frontmatter}

\title{Exact solution of a classical short-range spin model with a phase transition in one dimension: the Potts model with invisible states}

% else, use optional labels to link authors explicitly to addresses, as shown below:
\author[A,B,C]{Petro Sarkanych\fnref{myfootnote}}
\author[A,C]{Yurij Holovatch}
\author[B,C]{Ralph Kenna}
\address[A]{~Institute for Condensed Matter Physics, National Academy of Sciences of Ukraine, Lviv, Ukraine}
\address[B]{~Applied Mathematics Research Centre, Coventry University, Coventry, England}
\address[C]{~${\mathbb L}^4$ Collaboration \& Doctoral College for the Statistical Physics of Complex Systems, Leipzig-Lorraine-Lviv-Coventry, Europe}

\fntext[myfootnote]{sarkanyp@uni.coventry.ac.uk}

\begin{abstract}
    We present the exact solution of the 1D classical short-range Potts model with invisible states.
    Besides the   $q$ states of the ordinary Potts model, this possesses $r$ additional states which contribute
    to the entropy, but not to the interaction energy.  We determine the partition function, using the transfer-matrix
    method, in the general case of two ordering fields: $h_1$ acting on a visible state and $h_2$ on an  invisible
    state. We analyse its zeros in the complex-temperature plane  in the case that $h_1=0$. 
		When  ${\rm Im}\, h_2=0$ and $r\ge 0$, these zeros accumulate
    along a line that intersects the real temperature axis at the
    origin. This corresponds to the usual ``phase transition'' in a $1$D system.
    However, for   ${\rm Im}\, h_2\neq 0$ or $r<0$, the line of zeros intersects the positive part of the real
    temperature axis, which signals the  existence of a phase transition at non-zero temperature.
\end{abstract}


\end{frontmatter}

%\linenumbers

\section{\label{I}Introduction}

Few models in statistical physics  can be solved exactly and many of
those that can are in a single dimension  \cite{Baxter1982}. Knowing
the behaviour of a system in one dimension (1D) can help one
to understand and predict its behaviour in higher dimensions too.
Also some chemical compounds are effectively described by quasi-1D
models \cite{Azuma1994,Rice1993,Schulz1986,Ghulghazaryan07,Hovhannisyan09}. For these reasons, 1D
models are interesting from both theoretical and experimental points
of view.

In the 1920s Wilhelm Lenz  and Ernst Ising suggested and
investigated the first microscopic model of ferromagnetism
\cite{Lenz1920,Ising1925}. Theirs involved a one-dimensional lattice
occupied by classical spins which can only be  in one of two states:
up or down. In the short-range version, only nearest neighboring
spins are allowed to interact. Ising showed that this model has no
phase transition at any physically accessible (i.e., non-zero)
temperature $T$ \cite{Ising1925}. 
This served as the first and archetypal
 example of the absence of a finite-$T$ phase transition in
1D classical systems with short-range interactions \cite{Ising17}.
Later, in their famous book  \cite{Landau}, Landau and
Lifshitz gave a heuristic argument that for classical systems there
is no phase transition at non-zero temperature in 1D. The approach
is based on  separately evaluating contributions to the free
energy $F=E-TS$ coming from the interaction energy $E$ and the
entropy $S$.  For the  Ising model there are two
ground states in which all spins are either up or down. At zero
temperature the system is in one of these states. At finite
temperatures, domain walls separate regions of up- and down-spins.
Each domain wall ``costs'' energy (i.e., it increases the
interaction energy $E$). But in the 1D case this is
outweighed by the entropic contribution coming from the number of
ways of placing domain walls on the chain. See, e.g.,
Ref.\cite{Christensen2005} where it is explicitly shown that adding
domain walls reduces the free energy. Therefore, at any temperature
it is energetically favourable for domain walls to be inserted. This
means the system cannot become ordered and there is no phase
transition in such 1D systems.

van Hove proved the absence of a phase transition in one-dimensional
fluid-like systems of particles with  non-vanishing
incompressibility radius and a finite range of  forces
\cite{vanHove1950}. This was extended by Ruelle \cite{Ruelle1969} to
lattice models. These results are based on the Perron-Frobenius
theorem \cite{Gantmacher1959}. However, and as emphasised by Cuesta
and S{\'{a}}nchez, none of these theorems preclude the existence of
thermodynamic phase transitions in general 1D systems with
short-range interactions \cite{Cuesta2004}. Indeed Cuesta and
S{\'{a}}nchez gave examples of such models and Theodorakopoulos also
discussed how the no-go theorems might be circumvented
\cite{Theodorakopoulos2006}. Ref. \cite{Turban2016} maps the 1D 
multispin-interaction Ising model in a field onto a zero-field 
2D Ising model with nearest neighbours interaction. 
Additionally, quantum 1D models offer
further examples of systems which can undergo a phase transition at
non-zero temperatures because they are related to 2D classical
systems \cite{Suzuki1976}.

In the light of these enduring discussions, it is interesting to
investigate further circumstances in which classical systems might
exhibit a phase transition in 1D. In particular, we are interested
in a mechanism in which entropy production can be suppressed in
order to sidestep the conditions of the arguments and theorems
discussed above. To do so we consider the Potts model with invisible
states which was introduced a few years ago \cite{Tamura2010,
Tanaka2011} in order to explain some questions about the order of a
phase transition where $Z(3)$ symmetry is broken. It differs from
the ordinary Potts model \cite{Potts1952,Wu1982} by adding
non-interacting states; if a spin is in one such state, it is
``invisible'' to its neighbours. Originally, the corresponding
Hamiltonian was written in the form
\begin{equation}\label{1}
H=-\sum_{<i,j>}\delta _{S_i,S_j}\sum_{\alpha=1}^q \delta
_{S_i,\alpha}\delta _{S_j,\alpha},
\end{equation}
where $q$ and $r$ are the number of visible and invisible states of the Potts variable
\begin{equation}\label{2}
s_i=1,\ldots,q,q+1,\ldots,q+r \, .
\end{equation}
The first sum in (\ref{1}) is taken over all distinct pairs of interacting particles,
and the second sum requires
both of the interacting spins to be in the same visible state. From now and
on we will use the notation $(q,r)$-state Potts model for model with $q$ visible
and $r$ invisible states.

In the ordinary Potts model \cite{Potts1952}, the parameter $q\geq 2$ has 
been introduced as an integer denoting the number of (visible) states 
that a site can be in. 
However, it has been extended to other values too in order to describe
bond percolation ($q=1$), dilute spin glasses ($q=1/2$),  and gelation ($q=0$) \cite{Wu1982,Aharony1979,Lubensky1978}.
The parameter $q$ has been extended to complex values as well \cite{Glumac2002,Chang2013,Kim2001,Kim2003}.
In a similar way, although an initial interpretation of the parameter $r$ is the number of invisible states, 
we extend it here to non-integer and even negative values. 
As we will show below, the latter corresponds to removing entropy from the system
and will be key to inducing a phase transition.

Adding invisible states does not change the spectrum of the model, it only changes the degeneracy of energy levels
(the number of configurations with a given energy).
Even though invisible states do not change the interaction energy, since they change the entropy they affect the free energy.
As a consequences, for example,  an increase in the number $r$ of invisible states may cause a phase transition to change from  second to first order \cite{Tamura2010,Krasnytska2016}.
For example the $(2,30)$-state model on a square lattice undergoes the first order transition, while the ordinary
$(2,0)$-state Potts model (i.e. the Ising model) is an iconic example of a continuous transition.

The Potts model with invisible states describes a number of
models of physical interest. In particular, the $(1,r)$-state model
can be mapped to the Ising model in a temperature-dependent field
\cite{Badasyan2010}. The 1D $(1,r)$ case with nearest-neighbour
interactions  is equivalent to the Zimm-Bregg model for the
helix-coil transition \cite{Badasyan2010}. The multi-spin extension
of this model possesses a re-entrant phase transition and is in good
agreement with experimental observations for polymer transitions
\cite{Badasyan2011,Badasyan2012}. The $(2,r)$-state Potts model
without external fields is equivalent to the Blume--Emery--Grifiths
(BEG) model \cite{Tamura2010,Johnston2013,Ananikian2013} with a
temperature dependent external field. Furthermore, the general $q$
and $r$ case can be interpreted as a diluted Potts model
\cite{Tamura2010, Krasnytska2016}
\begin{equation}\label{3}
H^\prime=-\sum_{<i,j>}\delta _{\sigma_i,\sigma_j}(1-\delta_{\sigma_i,0}) - T\ln r\sum_i \delta_{\sigma_i,0} ,
\end{equation}
where $\sigma_i=0,1,\ldots,q$ is a new spin variable and all
invisible states are gathered into one the zeroth state $\sigma_i=0$.

In this paper we perform an analysis of the partition function zeros to obtain the exact solution
for the 1D Potts model with invisible states.
In achieving this, we add another exact result to the existing collection of exactly solved models in statistical mechanics.
As we will show below, although one-dimensional, the model manifests a second order
transition at non-zero temperature provided some unusual conditions are assumed.
We will discuss  regimes in which such behaviour is observed and a possible connection  with quantum systems.

The rest of the paper is organised as follows: in Section \ref{II} we present the
exact solution using the transfer-matrix method, then in Section \ref{III} the existence
of a phase transition at non-zero temperature is demonstrated using partition function zeros, and finally conclusions are given in Section \ref{IV}.

\section{\label{II} Exact solution of the Potts model with invisible states}
\quad
Let us consider the $(q,r)$-state Potts model on a chain consisting of $N$ spins  with only nearest-neighbour interactions subject to  two separate magnetic fields $h_1$ and $h_2$
acting on the first visible and the first invisible states respectively.
Imposing  periodic boundary conditions, the
Hamiltonian of such a system may be written as
\begin{eqnarray}
H_{(q,r)} & = & -\sum_{i}\sum_{\alpha=1}^q\delta_{s_i,\alpha}\delta_{\alpha,s_{i+1}}
\nonumber
 \\
 & & - h_1\sum_i \delta_{s_i,1} - h_2\sum_i\delta_{s_i,q+1} \, ,
\label{4}
\end{eqnarray}
where the sum over $i$ is taken over all sites of the chain.

We will use the transfer matrix formalism \cite{Katsura1972,Kim2000,Shrock1997} to
obtain the exact solution of the model (\ref{4}).
The Hamiltonian (\ref{4}) can be expressed as a sum of terms representing one bond each, so that $H_{(q,r)}=\sum_{i} H_{i}$ where
\begin{equation}\label{5}
H_{i}=-\sum_{\alpha=1}^q\delta_{s_i,\alpha}\delta_{\alpha,s_{i+1}}-h_1\delta_{s_i,1}-h_2\delta_{s_i,q+1} \, .
\end{equation}

Then the partition function can be transformed to
\begin{equation}\label{6}
Z=\sum_{s}\exp\left[{-\beta H_{(q,r)}}\right]=\sum_{s}\prod_{i}\exp\left( -\beta H_i\right)\, .
\end{equation}
where $\beta = 1/kT$ and $k$ is the Boltzmann constant. Now it is
easy to define the $(q+r) \times (q+r)$ square transfer matrix
 with elements
\begin{eqnarray}\label{7}
\hspace*{-9mm}
T_{ij}=\exp\left[{\beta\left({\delta_{s_i,s_{j}}\sum_{\alpha=1}^q\delta_{s_i,\alpha} + h_1\delta_{s_i,1} + h_2\delta_{s_i,q+1}}\right)}\right].
\hspace*{-2mm}
\end{eqnarray}
Let us denote
\begin{equation}\label{7a}
t=e^{-\beta}\, , \, z_1=e^{\beta h_1}\, , \, z_2=e^{\beta h_2}\, .
\end{equation}
With this notation, positive values of temperature $T$
correspond to $t$ ranging from zero to one, and the elements of the transfer
matrix can be written in the compact form: $T_{11}=z_1/t$;
$T_{ii}=t^{-1}$ for $1<i\leq q$; $T_{i1}=z_1$;
$T_{(q+1)i}=z_2$; and all remaining elements  equal to $1$.

Based on the transfer-matrix symmetry it is easy to show that it has five different eigenvalues.
The eigenvalue $\lambda_0=0$ is $(r-1)$ times degenerate because the final $r$ columns of the matrix are proportional.
Because $(q-1)$ elements of the main diagonal are equal to $t^{-1}$, choosing $\lambda=t^{-1}-1$ one can find $q-2$ linear independent eigenvectors.
This reduces the problem to the determination of three more eigenvalues.
They can be found using invariant permutations.
The above considerations lead to the equation for the three remaining eigenvalues:
\begin{eqnarray}\label{8}
\nonumber
 \hspace*{-15mm}( r+z_2-1-\lambda )
 [(\frac1t-1)z_1-\lambda]
 [(\frac1t-1)-\lambda]\\
 \hspace*{-5mm}- \lambda z_1[(\frac1t-1)-\lambda]
 -(q-1) \lambda [(\frac1t-1)z_1-\lambda]  =  0 \, .
\end{eqnarray}
This equation is of third order and can be solved exactly.
Since the partition function is
\cite{Katsura1972,Kim2000,Shrock1997}
\begin{equation}
\label{9}
Z=\lambda_1^N+\lambda_2^N+\lambda_3^N+\lambda_4^N,
\end{equation}
and all the $\lambda$'s have been found, the problem has been solved exactly.
In the thermodynamic limit $N\to \infty$ only the largest eigenvalue contributes
to the partition function.

It is worth mentioning  that the number of invisible states $r$ and the magnetic field $z_2$
appear as an additive combination   in Eq.(\ref{8});  the factor $r+z_2-1$ can be treated as
a temperature-dependent effective number of invisible states.

\section{\label{III} Partition function zeros}

Critical properties of any system can be  recovered from its
partition function. In the transfer-matrix approach, this is
expressed in terms of the transfer-matrix eigenvalues as in
Eq.(\ref{9}). Since all the $\lambda$'s can be found explicitly for
the Potts model with invisible states, its thermodynamic properties
can be found explicitly as well. Eq. (\ref{8}) is of third order,
making the exact solution cumbersome. Analysis of the
partition-function zeros in the complex temperature plain (Fisher
zeros) ameliorate this problem \cite{Fisher1965}. In the
thermodynamic limit they approach the real axis at the transition
point. For the 1D system, Fisher zeros can be obtained by equating
(at least) the two  eigenvalues which are largest by modulus
\cite{Fisher1980,Ghulghazaryan07,Hovhannisyan09}:
\begin{equation}
\label{10}
|\lambda_1|=|\lambda_2|>|\lambda_3|,\, |\lambda_4|\, .
\end{equation}
It is clear from this condition that $\lambda_2=e^{i\phi}\lambda_1$, where $\phi$ is a real phase factor.
Then the partition function (\ref{9})  attains the  form
\begin{equation}\label{11}
Z=|\lambda_1|^N\left[{1+e^{iN\phi}+\left(\frac{\lambda_3}{\lambda_1}\right)^N+\left(\frac{\lambda_4}{\lambda_1}\right)^N}\right].
\end{equation}
In the thermodynamic limit the third and the forth terms in the brackets are negligible and thus the phase $\phi$ is not arbitrary
but can take only the values $\phi=(2k-1)\pi/N$, where $k=1\dots N$.

\begin{figure}
    \begin{center}
        \includegraphics[width=0.4\paperwidth]{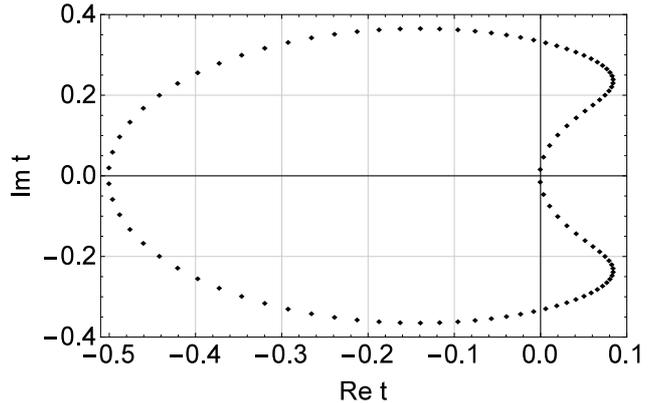}
        \caption{Fisher zeros for the (2,2)-state Potts model for  system size $N=100$ in the complex
        $t$--plane. Here $h_1= h_2=0$. The zeros approach the real $t$-axis at the point corresponding to the phase transition,
         namely at $t=0$ ($T=0$). \label{fig1}}
    \end{center}
\end{figure}

Consider first the case $h_1=h_2=0$ (i.e. $z_1=z_2=1$).
This  reduces the number of eigenvalues in (\ref{9}) and only three non-zero eigenvalues remain.
A typical plot of Fisher zeros that follows from Eqs. (\ref{10}) and (\ref{11})
is shown in Figure \ref{fig1}.
As one can see from the plot, the zeros approach the real axis at the point $t=0$, which corresponds to the zero-temperature phase transition.

The question that concerns us here is:
what are the circumstances in which the locus of Fisher zeros corresponds to a phase transition at non-zero temperature in 1D?
In the following, we discuss two cases where such behaviour is observed.

\subsection{The case $h_1=0$, $h_2 \in {C}$}
\label{3.1}

\begin{figure}
    \begin{center}
        \includegraphics[width=0.4\paperwidth, height=0.35\paperwidth]{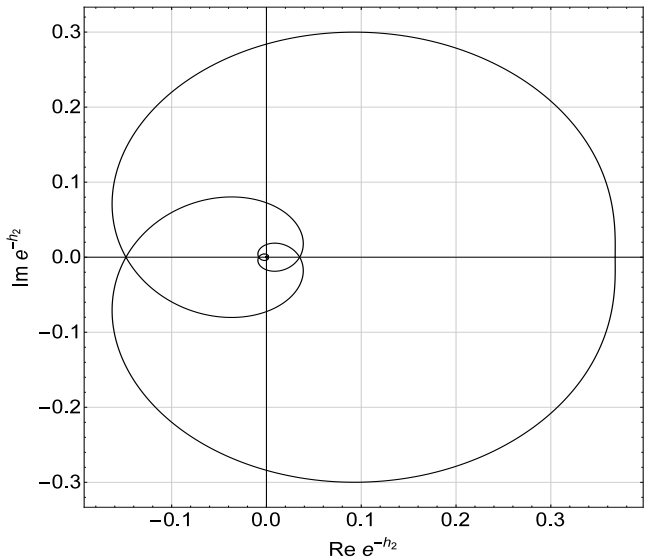}
        \caption{Values of $e^{-h_2}$, for which the phase transition in the (2,3)-state Potts model occurs at positive
        temperature. Each point of the plot corresponds to a certain physically accessible critical temperature.}\label{fig2}
    \end{center}
\end{figure}

We first demonstrate that when the field $h_1$ is zero, such behaviour can be achieved  by allowing field $h_2$ to be complex.
To this end we set the discriminant of Eq. (\ref{8}) equal to zero for positive temperature and $h_1=0, h_2 \in {C}$.
This leads to a relation between the critical temperature and field $h_2$, namely
\begin{equation}\label{12}
z_2=t^{-1}-q-r \pm 2 i \sqrt{q (t^{-1}-1)}\, .
\end{equation}
In this equation, the temperature resides in $t$ and in $z_2=t^{-h_2}$.
For a given (real) value of $t$ between $0$ and $1$ [$T \in (0,\infty)$], we numerically obtain
complex values for $h_2=\rm{Re}\,h_2 + i {\rm{Im}} \, h_2$.
We plot these values in the form $e^{-h_2}$ in Fig.\ref{fig2} for the example where $q=2$, $r=3$.
Every strictly complex point on the plot corresponds to a positive real value of $T$.

The resulting plot has a form of
the double spiral, two branches of the spiral correspond to two complex conjugate values of the field.

Obviously, the variable $h_2$ does not have a direct, real physical interpretation; it is a complex field acting on invisible states.
As expected, the finite-temperature phase transition does not occur in the 1D  system under real physical conditions \cite{Landau,Ruelle1969}.
However, it was recently suggested that the complex magnetic field in a classical system can be mapped onto decoherence time \cite{Wei2012} in its quantum counterpart \cite{Wei2012}.
Of course, in quantum systems finite-temperature phase transitions are allowed, even in 1D \cite{Suzuki1976}.
Effectively, rendering the field $h_2$ complex, as was done above, adds additional degrees of freedom to the system.
Similarly, an increase of the number of degrees of freedom occurs when one passes from a classical to a quantum description.
In this sense, the phase transition described above, although occurring in an unphysical region of parameter space, may have a
counterpart in the quantum world. Moreover, the external field $h_2$ can be tuned to change the  system's behaviour from quantum
into classical and vice versa.

\subsection{The case $h_1=0$, $h_2=0$ and $r<0$}
\label{3.2}
Another way to invoke a phase transition
is to relax the condition of positivity on the number of invisible states $r$.
As we noted before, the free energy of the one-dimensional system with two mixed
phases decreases with  increasing the number of  domain walls, so that an ordering
phase transition is not possible for finite temperature \cite{Landau}.
As a way to circumvent this restriction one can try to introduce a mechanism that
leads to entropy decrease and therefore to the free energy increase.
The calculations displayed below are based on the obvious observation that if adding
invisible states increases the entropy, a negative number of invisible states decreases it.

\begin{figure}
    \begin{center}
        \includegraphics[width=0.4\paperwidth]{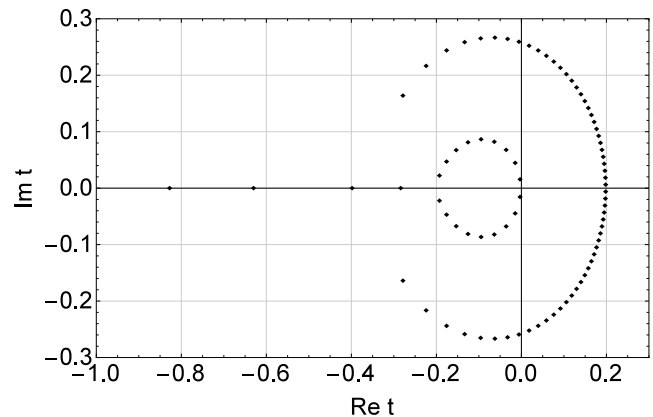}
        \caption{
            Fisher zeros for the (2,-6)-state Potts model for system size $N=100$ in the complex   $t$--plane. The Fisher zeros cross the real axis at the point $t=0.2$ corresponding to a physical temperature of $T=0.621$. \label{fig3}}
    \end{center}
\end{figure}

To demonstrate the behaviour of the Potts model with a negative number of invisible states,
in Figure \ref{fig3} we display Fisher zeros for the (2,-6)-state Potts model.\footnote{We do not show in the plot some points in
the region $t>1$ that correspond to negative temperatures.}

As one can see from the figure, the locus of Fisher zeros in this case intersects the real axis at a positive value of the temperature $t$.
The equivalent representation of the Potts model with invisible states through Eq. (\ref{3}) indicates that the chemical potential is $\mu=-T\log r$.
Therefore  the negative number of invisible states is equivalent to a model with complex chemical potential.
Again, via the aforementioned relation  between the complex external field and decoherence time \cite{Wei2012}, this gives a connection to the behaviour of quantum systems.

Although both methods introduced above can shift the critical point from zero to positive values, the order of the phase transition and the values of the critical exponents remain the same as for the 1D  Ising model, namely $\nu=1,\, \alpha=1,\,\eta=1,\,\gamma=1,\,\mu=0,\,\beta=0,\,\delta=\infty$ \cite{Baxter1982}.

\section{\label{IV} Conclusions}

One dimensional  models of statistical physics are interesting from both theoretical and experimental points
of view, often opening the way to develop methods for solving more complex problems.
In this paper we exactly solved the Potts model with invisible states with two ordering fields in 1D  using the
transfer-matrix method.

Various authors  gave arguments and theorems delivering the absence of positive-temperature  phase transitions
in classical 1D  systems \cite{Landau,vanHove1950,Ruelle1969,Cuesta2004}. 
These can only exhibit properties characteristic of second-order phase transitions at zero temperature \cite{Baxter1982}.
Here we introduced two methods to circumvent these arguments and theorems to induce phase transitions at 
physically accessible temperatures in a 1D short-range system.

The first method involves placing the system into a complex external field acting on an invisible state.
This method allows one to achieve any critical temperature by tuning of the field.
In  recent publications  it was shown that a complex external field for a  classical system can be mapped to the decoherence
time of a quantum counterpart system \cite{Wei2012}.
Thus allowing the external field to be complex might be interpreted as giving a classical system certain quantum properties.

The second method we have introduced here involves  using  negative numbers of invisible states $r$.
This way may be interpreted as introducing some  ordering factor to the system, reducing its entropy.
An alternative interpretation is in terms of a complex chemical potential.

The reason why rigorous theorem proved in \cite{Ruelle1969} doesn't
hold here lies in the fact, that the proof was made using
Perron-Frobenius theorem, which is applicable only for matrices of
real positive values. Once either the external field or the chemical
potential becomes complex, this statement, in general, is not true.
Regarding the arguments given by Landau and Lifshitz, it concerns
systems which tend to the state with free energy minimum. For
complex values of parameters, energy becomes complex as well,
and for complex numbers minimum is not defined.

These results offer further counter-examples to the often asserted  general statement that there can be no positive-temperature phase transitions in 1D systems with short-range interactions. 
Indeed, and as convincingly pointed out in Ref.\cite{Cuesta2004}, although 1D systems are very thoroughly researched, 
they are still capable of yielding exciting new physics.

{\bf Acknowledgements:}
We would like to thank Nerses Ananikian and  Vahan Hovhannisyan for fruitful discussions. 
This work was supported in part by FP7 EU IRSES projects No. 612707 ``Dynamics of and in Complex Systems'' and
No. 612669 ``Structure and Evolution of     Complex Systems with
Applications in Physics and Life Sciences''.


\begin{thebibliography}{99}


\bibitem{Baxter1982}
Baxter, R. J. (1982). Exactly solved models in statistical physics. Academic, New York.


\bibitem{Azuma1994}
Azuma, M., Hiroi, Z., Takano, M., Ishida, K., and Kitaoka, Y. (1994).
%Observation of a Spin Gap in $Sr Cu_2O_3$ Comprising Spin-1/2 Quasi-1D Two-Leg Ladders.
Physical review letters, {\bf{73}}, 3463.

\bibitem{Rice1993}
Rice, T. M., Gopalan, S., and Sigrist, M. (1993).
%Superconductivity, spin gaps and Luttinger liquids in a class of cuprates.
EPL (Europhysics Letters), {\bf 23}, 445.


\bibitem{Schulz1986}
Schulz, H. J. (1986).
%Phase diagrams and correlation exponents for quantum spin chains of arbitrary spin quantum number.
Physical Review B, {\bf 34}, 6372.

\bibitem{Ghulghazaryan07}
Ghulghazaryan, R. G., Sargsyan, K. G., and Ananikian, N. S. (2007).
Physical Review E, {\bf 76}, 021104.

\bibitem{Hovhannisyan09}
Hovhannisyan, V. V., Ghulghazaryan, R. G., and Ananikian, N. S.
(2009). Physica A: Statistical Mechanics and its Applications, {\bf
	388}, 1479. 

\bibitem{Lenz1920}
Lenz, W. (1920).
%Beitrag zum Verst\"{a}ndnis der magnetischen Erscheinungen in festen K\"{o}rpern.
Zeitschrift f{\"u}r Physik, {\bf  21}, 613.


\bibitem{Ising1925}
Ising, E. (1925).
%Beitrag zur theorie des ferromagnetismus.
Zeitschrift f{\"u}r Physik, {\bf 31}, 253.

\bibitem{Ising17}
Ising T., Folk R., Kenna R., Berche B., and Holovatch Yu. (2017)
preprint arXiv:1706.01764.

\bibitem{Landau}
Landau, L. D., and Lifshitz, E. M. (1969). Statistical Physics: V. 5: Course of Theoretical Physics. Pergamon press.


\bibitem{Christensen2005}
Christensen, K., and Moloney, N. R. (2005). Complexity and criticality. Imperial College Press.


\bibitem{vanHove1950}
van Hove, L. (1950).
%Sur l'int{\'e}grale de configuration pour les syst{\`e}mes de particules {\`a} une dimension.
Physica, 16(2), 137.

\bibitem{Ruelle1969}
Ruelle, D. (1969). Rigorous results in statistical mechanics. Lectures in Theoretical Physics.


\bibitem{Gantmacher1959}
Gantmacher, F. R. (1959). Matrix Theory. Chelsea, New York.


\bibitem{Cuesta2004}
Cuesta, J. A., and S\'{a}nchez, A. (2004).
%General non-existence theorem for phase transitions in one-dimensional systems with short range interactions, and physical examples of such transitions.
Journal of statistical physics, {\bf 115}, 869.

\bibitem{Theodorakopoulos2006}
Theodorakopoulos, N. (2006).
%Phase transitions in one dimension: Are they all driven by domain walls?.
Physica D: Nonlinear Phenomena, {\bf 216}, 185.

\bibitem{Turban2016}
Turban, L. (2016). 
%One-dimensional Ising model with multispin interactions. 
Journal of Physics A: Mathematical and Theoretical, {\bf 49}, 355002.

\bibitem{Suzuki1976}
Suzuki, M. (1976).
%Relationship between d-dimensional quantal spin systems and (d+ 1)-dimensional Ising systems: Equivalence, critical exponents and systematic approximants of the partition function and spin correlations.
Progress of theoretical physics, {\bf 56}, 1454.

\bibitem{Tamura2010}
Tamura, R., Tanaka, S., and Kawashima, N. (2010).
%Phase transition in Potts model with invisible states.
Progress of theoretical physics, {\bf 124}, 381.

\bibitem{Tanaka2011}
Tanaka, S., and Tamura, R. (2011).
%Dynamical properties of Potts model with invisible states.
Journal of Physics: Conference Series, {\bf 320}, 012025.

\bibitem{Potts1952}
Potts, R. B. (1952).
%Some generalized order-disorder transformations.
Mathematical proceedings of the Cambridge philosophical society, {\bf 48}, 106.


\bibitem{Wu1982}
Wu, F. Y. (1982).
%The Potts model.
Reviews of modern physics, {\bf 54}, 235.


\bibitem{Aharony1979}
Aharony, A., and Pfeuty, P. (1979).
%Dilute spin glasses at zero temperature and the 1/2-state Potts model.
Journal of Physics C: Solid State Physics, {\bf 12}, L125.

\bibitem{Lubensky1978}
Lubensky, T. C., and Isaacson, J. (1978).
%Field theory for the statistics of branched polymers, gelation, and vulcanization.
Physical Review Letters, {\bf 41}, 829.

\bibitem{Glumac2002}
Glumac, Z., and Uzelac, K. (2002).
%Complex-q zeros of the partition function of the Potts model with long-range interactions.
Physica A: Statistical Mechanics and its Applications, {\bf 310}, 91.

\bibitem{Chang2013}
Chang, S. C., and Shrock, R. (2013).
%Zeros of the Potts model partition function on Sierpinski graphs.
Physics Letters A, {\bf 377}, 671.

\bibitem{Kim2001}
Kim, S. Y., and Creswick, R. J. (2001).
%Density of states, Potts zeros, and Fisher zeros of the Q-state Potts model for continuous Q.
Physical Review E, {\bf 63}, 066107.

\bibitem{Kim2003}
Kim, S. Y. (2004).
%Fisher zeros and Potts zeros of the Q-state Potts model for nonzero external magnetic field.
Journal of the Korean Physical Society, {\bf 45}, 302.


\bibitem{Krasnytska2016}
Krasnytska, M., Sarkanych, P., Berche, B., Holovatch, Y., and Kenna, R. (2016).
%Marginal dimensions of the Potts model with invisible states.
Journal of Physics A: Mathematical and Theoretical, {\bf 49}, 255001.

\bibitem{Badasyan2010}
Badasyan, A. V., Giacometti, A., Mamasakhlisov, Y. S., Morozov, V. F., and Benight, A. S. (2010).
%Microscopic formulation of the Zimm-Bragg model for the helix-coil transition.
Physical Review E, {\bf 81}, 021921.


\bibitem{Badasyan2011}
Badasyan, A. V., Tonoyan, S. A., Mamasakhlisov, Y. S., Giacometti, A., Benight, A. S., and Morozov, V. F. (2011).
%Competition for hydrogen-bond formation in the helix-coil transition and protein folding.
Physical Review E, {\bf 83}, 051903.


\bibitem{Badasyan2012}
Badasyan, A., Tonoyan, S., Giacometti, A., Podgornik, R., Parsegian, V. A., Mamasakhlisov, Y., and Morozov, V. (2012).
%Osmotic pressure induced coupling between cooperativity and stability of a helix-coil transition.
Physical review letters, {\bf 109}, 068101.


\bibitem{Johnston2013}
Johnston, D. A., and Ranasinghe, R. P. K. C. M. (2013).
%Potts models with (17) invisible states on thin graphs.
Journal of Physics A: Mathematical and Theoretical, {\bf 46}, 225001.


\bibitem{Ananikian2013}
Ananikian, N., Izmailyan, N. S., Johnston, D. A., Kenna, R., and Ranasinghe, R. P. K. C. M. (2013).
%Potts models with invisible states on general Bethe lattices.
Journal of Physics A: Mathematical and Theoretical, {\bf 46}, 385002.


\bibitem{Katsura1972}
Katsura, S., and Ohminami, M. (1972).
%Distribution of zeros of the partition function for the one dimensional Ising models.
Journal of Physics A: General Physics, {\bf 5}, 95.


\bibitem{Kim2000}
Kim, S. Y., and Creswick, R. J. (2000).
%Exact results for the zeros of the partition function of the Potts model on finite lattices.
Physica A: Statistical Mechanics and its Applications, {\bf 281}, 252.

\bibitem{Shrock1997}
Shrock, R., and Tsai, S. H. (1997).
%Complex-temperature phase diagrams of one-dimensional spin models with next-nearest-neighbor couplings.
Physical Review E, {\bf 55}, 5184.


\bibitem{Fisher1965}
Fisher, M. E. (1964). Lectures in Theoretical Physics: Volume VII. University of Colorado, Boulder.


\bibitem{Fisher1980}
Fisher, M. E. (1980).
%Yang-Lee edge behavior in one-dimensional systems.
Progress of Theoretical Physics Supplement, {\bf 69}, 14.

\bibitem{Wei2012}
Wei, B. B., and Liu, R. B. (2012).
%Lee-Yang zeros and critical times in decoherence of a probe spin coupled to a bath.
Physical review letters, {\bf 109}, 185701;
Peng, X., Zhou, H., Wei, B. B., Cui, J., Du, J., and Liu, R. B. (2015).
%Experimental observation of Lee-Yang zeros.
Physical review letters, {\bf 114}, 010601.


%\bibitem{Wang}
%X. Wang,
%Physical Examples of Phase Transition in
%One-Dimensional Systems with Short Range
%interaction,
%\url{http://guava.physics.uiuc.edu/~nigel/courses/563/Essays_2012/PDF/wang.pdf}

\end{thebibliography}
\end{document}